# Towards sustainable space research in France

Alexandre Santerne[1,*], Héloïse Meheut[2], Didier Barret[3], Olivier Berné[3], Etienne Berthier[4], Agnès Ducharne[5,6], Jürgen Knödlseder[3], Aurélie Marchaudon[3], Thierry Pellarin[7], Aymeric Spiga[8], Peter Wolf[9]

[1] Aix Marseille Univ, CNRS, CNES, LAM, Marseille, France
[2] Université Côte d'Azur, Observatoire de la Côte d'Azur, CNRS, Laboratoire Lagrange, France
[3] Institut de Recherche en Astrophysique et Planétologie, CNRS, Université Toulouse 3, CNES, 9 Avenue du Colonel Roche, Toulouse, 31028, France
[4] Université de Toulouse, LEGOS (CNES/CNRS/IRD/UT3), Toulouse, France
[5] Sorbonne Université, CNRS, EPHE, UMR 7619 METIS, Paris, France
[6] Institut Pierre Simon Laplace (IPSL), Sorbonne Université, CNRS, Paris, France
[7] Université Grenoble Alpes, CNRS, INRAE, IRD, Grenoble INP, IGE, Grenoble, France
[8] Sorbonne Université, Laboratoire de Météorologie Dynamique, Paris, France
[9] LTE, Observatoire de Paris-PSL, CNRS, Sorbonne Université, LNE, 61 Av. de l'Observatoire, 75014 Paris, France
* corresponding author (alexandre.santerne@lam.fr)

Ten years ago, 195 countries signed the so-called Paris Agreement that aims at limiting global warming within 2 degrees compared to pre-industrial time. According to the International Panel on Climate Change (IPCC), this objective can be only reached if anthropocentric greenhouse gas (GHG) emissions decrease to net zero by 2050 [1]. To achieve this challenge, all human activities, including scientific research, must become sustainable. In France, awareness on this topic is shared by a large fraction of the academic community [2]. One third of the French research laboratories already started evaluating their GHG emissions, using the public tool developed by Labos1point5 [3,4]. In 2022, the ethics committee of the French national scientific research center (CNRS) stressed that "*the environmental impact of research should be considered as part of research ethics, in the same way as respect of human beings and experimental animals*".

In this context, research performed thanks to space-based instrumentations is facing formidable challenges. They make an unequivocal contribution to society by benefiting from the unique conditions of space. However, space missions have a large carbon footprint [5,6,7], notably because they need to be launched by rockets that are difficult to decarbonise. Rocket's debris and satellite re-entries are also impacting the stratosphere [8] and the ozone layer [9]. Nevertheless, in January 2024, CNES (the French space agency), together with 15 other French national research organisations, committed to "*contribute, through their research activities and changes in their operations, to meet the challenges of the ecological transition for sustainability*". In particular, they decided to be "*an example in the application of the objectives of [France's] ecological plan*". The European Space Agency (ESA) also committed to "*reducing [their] greenhouse gas emissions by 46% by 2030 compared to the 2019 baseline*".

As part of CNES 2024's quinquennial scientific roadmap, a focus was placed on the sustainability of scientific research from space. For the first time, a specific group was mandated to make recommendations on how to "*decrease the environmental footprint of space science activities*". An assessment of the environmental footprint of a dozen typical space missions has yielded a set of recommendations. They address the various aspects of space science, from the funding of research teams to the French participation in international projects. These recommendations can be summarised as follow:
- quantify the carbon footprint of the national activities in space science on an annual basis;
- adopt a trajectory to decrease the overall GHG emissions of these activities, with an objective of -7 % per year;
- set a carbon budget for the GHG emissions of all stakeholders (researchers, industries, partners, etc…) ;
- define and select the space missions, as well as research and development activities, within those limits ;
And more generally one should:
- raise awareness of the environmental transition among all the actors of the national space sciences, including industrial partners and users of space data ;
- optimise the existing facilities, by increasing their lifetime, maximise the scientific use of data, share the ground equipments, as well as re-use, recycle, and eco-design whenever possible;
- promote cooperation, sharing of data and reduce competition at the international level;
- consider environmental impacts among the top criteria for scientific space policy, and anticipate the need for resilience.

These recommendations fully align with a white paper to CNES co-signed by more than 260 French scientists from both astrophysics and Earth sciences. To follow a -7% / year decrease of GHG emissions, space science should rely on two complementary methods: decarbonisation and frugality [10]. The former relies on developing and massively deploying new techniques of eco-design while reducing the GHG emissions of existing facilities, including the launch segment. Yet, there is no guarantee that decarbonisation will be sufficient in view of the urgency of the situation. The first actions are straightforward to implement [11] but it will likely take a lot of time to develop low-carbon technologies in the space industry to decrease GHG emissions by ~83%, such as to reach net-zero emissions by 2050. Moreover, the optimisation of space missions to decrease their carbon footprint will likely be associated with a rebound effect [12] that could even increase the overall GHG emissions of space research.

On the other hand, frugality consists in either decreasing the number or reducing the scope of new space missions. The former will be efficient to meet with the trajectory but comes with the risk of losing the unique know-how needed to develop space instruments. The next generations might also need new instruments and not only capitalise on the unexploited data from missions developed by previous generation researchers. One straightforward way to proceed would be to avoid competition while increasing the cooperation between countries. The study performed for the CNES roadmap found that small satellites have a larger footprint, relative to their weight or cost, than big ones. It is therefore better to prioritise a few large new space missions that will revolutionise scientific knowledge rather than developing many small and incremental projects. In view of these factors, the trajectory consistent with the Paris Agreement is likely to be achieved by combining both decarbonisation and frugality.

The report from the CNES roadmap also discusses the longer term strategies. Since these activities rely on public funds, they are deeply dependent on political decisions and the evolution of society, which are severely impacted by the climate crisis [13]. The report thus recommends anticipating these major societal changes, all the more so as space projects span several decades. We collectively need to make sure that the decisions taken now are compatible with the inevitable changes that will occur in society. This adaptation needs to consider all possible scenarios, whether global warming is kept within +2ºC or continues to rise in a business-as-usual scenario.

The space sector used to have a culture of risk management to avoid failure. It should now include the risks associated with the climate crisis and environmental effects. The space sector is a major innovation provider and should now focus it towards sustainable scientific space activities, using a combination of decarbonisation and frugality. This appears to be the most efficient method for pursuing space science over time, preserving it to the next generations. With the recommendations of this roadmap, CNES has the potential to move towards sustainable space research in France.

**Conflict of interest:**
The authors declare no competing interests

**References:**
[1] IPCC, 2018: Global warming of 1.5°C. An IPCC Special Report on the impacts of global warming of 1.5°C above pre-industrial levels and related global greenhouse gas emission pathways, in the context of strengthening the global response to the threat of climate change, sustainable development, and efforts to eradicate poverty [V. Masson-Delmotte, P. Zhai, H. O. Pörtner, D. Roberts, J. Skea, P.R. Shukla, A. Pirani, W. Moufouma-Okia, C. Péan, R. Pidcock, S. Connors, J. B. R. Matthews, Y. Chen, X. Zhou, M. I. Gomis, E. Lonnoy, T. Maycock, M. Tignor, T. Waterfield (eds.)].
[2] Blanchard, M., Bouchet-Valat, M., Cartron, D., Greffion, J., & Gros, J. (2022). Concerned yet polluting: A survey on French research personnel and climate change. PLOS Climate, 1(9), e0000070.
[3] Mariette, J., Blanchard, O., Berné, O., Aumont, O., Carrey, J., Ligozat, A., ... & Ben-Ari, T. (2022). An open-source tool to assess the carbon footprint of research. Environmental Research: Infrastructure and Sustainability, 2(3), 035008.
[4] Ben-Ari, T. (2023). How research can steer academia towards a low-carbon future. Nature Reviews Physics, 5(10), 551-552.
[5] Knödlseder, J., Brau-Nogué, S., Coriat, M., Garnier, P., Hughes, A., Martin, P., & Tibaldo, L. (2022). Estimate of the carbon footprint of astronomical research infrastructures. Nature Astronomy, 6(4), 503-513.
[6] Knödlseder, J., Coriat, M., Garnier, P., & Hughes, A. (2024). Scenarios of future annual carbon footprints of astronomical research infrastructures. Nature Astronomy, 8(11), 1478-1486.
[7] Marc, O., Barret, M., Biancamaria, S., Dassas, K., Firmin, A., Gandois, L., ... & Toublanc, F. (2024). Comprehensive carbon footprint of Earth, environmental and space science laboratories: Implications for sustainable scientific practice. *PLOS Sustainability and Transformation*, *3*(10), e0000135.
[8] Murphy, D. M., Abou-Ghanem, M., Cziczo, D. J., Froyd, K. D., Jacquot, J., Lawler, M. J., ... & Shen, X. (2023). Metals from spacecraft reentry in stratospheric aerosol particles. Proceedings of the National Academy of Sciences, 120(43), e2313374120.

[9] Ferreira, J. P., Huang, Z., Nomura, K. I., & Wang, J. (2024). Potential ozone depletion from satellite demise during atmospheric reentry in the era of mega-constellations. Geophysical Research Letters, 51(11), e2024GL109280.

[10] Saint-Martin, A. (2023). Une astronautique «frugale»?. Histoire et croissance d'un mantra. Socio. La nouvelle revue des sciences sociales, (17), 187-213.

[11] Barret, D., Albouys, V., Knödlseder, J., Loizillon, X., D'Andrea, M., Ardellier, F., ... & Webb, N. (2024). Life cycle assessment of the Athena X-ray integral field unit. Experimental Astronomy, 57(3), 19.

[12] Schaefer, S., & Wickert, C. (2015). The efficiency paradox in organization and management theory. In Academy of Management Proceedings (Vol. 2015, No. 1, p. 10958). Briarcliff Manor, NY 10510: Academy of Management.

[13] IPCC, 2022: Climate Change 2022: Impacts, Adaptation, and Vulnerability. Contribution of Working Group II to the Sixth Assessment Report of the Intergovernmental Panel on Climate Change [H.-O. Pörtner, D.C. Roberts, M. Tignor, E.S. Poloczanska, K. Mintenbeck, A. Alegría, M. Craig, S. Langsdorf, S. Löschke, V. Möller, A. Okem, B. Rama (eds.)]. Cambridge University Press. Cambridge University Press, Cambridge, UK and New York, NY, USA, 3056 pp., doi:10.1017/9781009325844.